\documentclass[pra, two column]{revtex4}%
\usepackage{amsfonts}
\usepackage{amsmath}
\usepackage{amssymb}
\usepackage{graphicx}%
\providecommand{\U}[1]{\protect\rule{.1in}{.1in}}
\begin{document}
\title{Nonlinear Floquet dynamics of spinor condensates in an optical cavity:
Cavity-amplified parametric resonance}
\author{Zheng-Chun Li$^{1}$, Qi-Hui Jiang$^{1}$, Zhihao Lan$^{2}$, Weiping
Zhang$^{3,4}$, and Lu Zhou$^{1,4}$\footnote{Corresponding author:
lzhou@phy.ecnu.edu.cn }}
\affiliation{$^{1}$State Key laboratory of Precision Spectroscopy, Department of Physics,
School of Physics and Electronic Science, East China Normal University,
Shanghai 200241, China}
\affiliation{$^{2}$Department of Electronic and Electrical Engineering, University College
London, Torrington Place, London, WC1E 7JE, United Kingdom}
\affiliation{$^{3}$Department of Physics and Astronomy, Shanghai Jiaotong University and
Tsung-Dao Lee Institute, Shanghai 200240, China}
\affiliation{$^{4}$Collaborative Innovation Center of Extreme Optics, Shanxi University,
Taiyuan, Shanxi 030006, China}

\begin{abstract}
We investigate Floquet dynamics of a cavity-spinor Bose-Einstein condensate
coupling system via periodic modulation of the cavity pump laser. Parametric
resonances are predicted and we show that due to cavity feedback-induced
nonlinearity the spin oscillation can be amplified to all orders of resonance,
thus facilitating its detection. Real-time observation on Floquet dynamics via
cavity output is also discussed.

\end{abstract}
\maketitle

%\pacs{}

\section{introduction}

\label{sec_introduction}

As one promising scheme of implementing quantum engineering, Floquet dynamics
have been widely studied in many quantum systems
\cite{goldmanPRX2014,shirley,salzman,dietz and holthaus}. The interest lies in
that one can substantially modify the long-time dynamical properties of a
quantum system via driving it with a short-time period as well as its
potential in realizing novel new quantum device, which was demonstrated in
tremendous experimental and theoretical works such as matter wave jet
\cite{chinNature2017,fengScience2019}, Floquet-Bloch bands
\cite{holthausJPB2016,weldPRL2019}, Bloch oscillation in a two-band model
\cite{plotz}, quantum ratchets
\cite{smirnovPRL2008,grifoniPRL2002,creffieldPRL2007,lundhPRL2005,chienPRA2013,wimberger,mark}%
, driven optical lattices \cite{eckardtRMP2017,fanPRA2019,kolovsky}, kicked
rotor \cite{rotor}, Floquet time crystal \cite{elsePRL2016,huangPRL2018} and
monopole magnetic field \cite{zhouPRL2018}.

Very recent experiments have demonstrated Floquet dynamics in spinor $^{87}$Rb
Bose-Einstein condensate (BEC), with the emphasis on spin oscillation
\cite{chapmanNC2016,gerbier2018} and quantum walk in momentum space
\cite{gil}, respectively. Experimental realization of spinor BEC has opened up
a new research direction of cold atom physics \cite{stengerNature1998}, in
which superfluidity and magnetism are simultaneously achieved. In spinor BEC
the spin-dependent collision interactions \cite{spinor condensate} allow for
the population exchange among hyperfine spin states and give rise to coherent
spin-mixing dynamics
\cite{uedaPR2012,hanPRL1998,chapmanNP2005,lettPRL2007,zhangPRA2005,wideraPRL2005,kronjagerPRL2006,gerbierPRA2012R,shinPRL2016}%
. In principle spin-mixing is Josephson-like effect that takes place in
internal degrees of freedom of atomic spin as compared with that in external
degrees of freedom such as a BEC in a double-well potential, for which the
Floquet dynamics can be studied via periodic modulation of the barrier height
(Josephson coupling) or the difference between the well depths
\cite{haiPRA2009,haroutyunyanPRA2004,bigelowPRA2005,eckardtPRL2005,wangPRA2006,kuangPRA2000,ashhabPRA2007}%
. Similar to that, in the two experiments \cite{chapmanNC2016,gerbier2018}
magnetic field plays an important role, which modifies the relative energy
among spin states via the quadratic Zeeman effect. Parametric resonance (or
Shapiro resonance) and spin oscillation have been observed via applying biased
magnetic field.

On the other hand besides the magnetic field, recent years have witnessed
growing interest in mediating atomic dynamics via the coupling of BEC to an
optical cavity \cite{cold atom and cavity review}. With the aid of cavity
light field, researchers have successfully implemented photon-mediated
spin-exchange interactions \cite{norciaScience2018,davisPRL2019}, formation of
spin texture \cite{donnerPRL2018} and spinor self-ordering \cite{levPRL2018}.
Cavity-induced superfluid-Mott insulator transition
\cite{larsonPRL2008,zhouPRA2013}, cavity backaction-driven atom transport
\cite{goldwinPRL2014} and BECs with cavity-mediated spin-orbit coupling are
also reported \cite{dongPRA2014R,dengPRL2014}. In these works cavity feedback
plays an important role.

In this work, by considering the fact that effective quadratic Zeeman effect
can be generated by a strong off-resonant laser field \cite{santosPRA2007}, we
propose an experimentally feasible scheme to realize cavity-driven Floquet
dynamics in spinor BECs. An interesting problem in this setup is that the
Floquet dynamics and the modulating parameter will become mutually dependent
through the cavity feedback. As compared with previous theoretical works
\cite{cosmePRL2018,zhangPRL2018} in which cavity drives the external
centre-of-mass motion of the BECs, here we will look into the problem of what
will take place in the "internal" Floquet dynamics of a spinor BEC driven by
the cavity light field.

The article is organized as follows: In Sec. \ref{sec_model} we present our
model and the effective Hamiltonian is derived for the driven system on
resonance. Sec. \ref{sec_amplify} is devoted to the discussion of how the
Floquet dynamics are affected by the cavity-induced nonlinearity. The
possibility of performing real-time observation of the Floquet dynamics in the
present system is explored in Sec. \ref{sec_measure}. Finally we conclude in
Sec. \ref{sec_conclusion}.

\section{model}

\label{sec_model}

\begin{figure}[h]
\includegraphics[width=8cm]{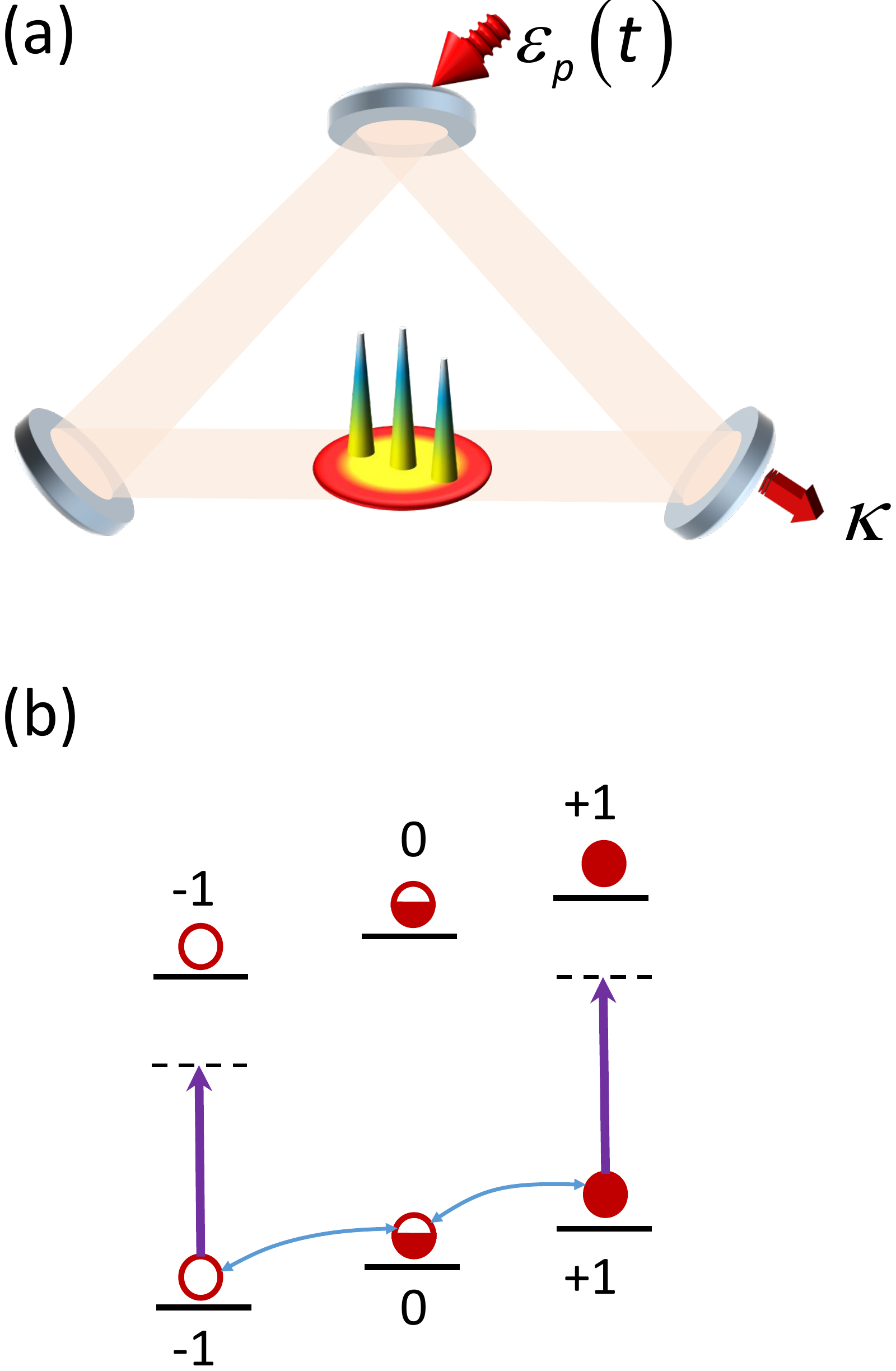}\caption{{\protect\footnotesize Schematic
diagram for generating cavity-amplified parametric resonance. (a) An }$F=1$
{\protect\footnotesize spinor condensate is trapped inside a ring cavity. The
cavity is coherently driven by an external laser with time-dependent amplitude
}$\varepsilon_{p}\left(  t\right)  ${\protect\footnotesize \ and decays with a
rate }$\kappa${\protect\footnotesize . (b) The cavity field is }$\pi
${\protect\footnotesize -polarized and is dispersively coupled to the atomic
system. In the meanwhile the spin-dependent collisions will lead to population
transfer among the three spin components.}}%
\label{fig_scheme}%
\end{figure}

We consider the following model depicted in Fig. \ref{fig_scheme}: A spinor
BEC of $^{87}$Rb atoms with hyperfine spin $F_{g}=1$ confined in an optical
dipole trap is placed inside a unidirectional ring cavity. The intracavity
mode is driven by a coherent laser field with frequency $\omega_{p}$ and
time-dependent amplitude $\varepsilon_{p}\left(  t\right)  $, which we assume%
\begin{equation}
\varepsilon_{p}\left(  t\right)  =\varepsilon_{0}\left[  1+f_{0}\sin\left(
\omega_{m}t\right)  \Theta\left(  t\right)  \right]  , \label{eq_drive}%
\end{equation}
with $\Theta\left(  t\right)  $ the Heaviside step function\ implying that a
sinusoidal modulation around a bias value $\varepsilon_{0}$ is activated at
$t=0$.\ The cavity mode is described by an annihilation operator $\hat{a}$,
which is $\pi$-polarized and characterized by a frequency $\omega_{c}$ and a
decay rate $\kappa$. Furthermore we assume that $\omega_{c}$ is detuned away
from the $F_{g}=1\longleftrightarrow F_{e}=1$ atomic transition such that the
atom-photon interaction is essentially of dispersive nature. The transition
selection rule allows states $\left\vert F_{g}=1,m_{g}=\pm1\right\rangle $ to
be coupled to the corresponding states in the excited manifold with the same
magnetic quantum numbers $\left\vert F_{e}=1,m_{e}=\pm1\right\rangle $ while
it forbids state $\left\vert F_{g}=1,m_{g}=0\right\rangle $ to make dipole
transitions to any excited states. The resulting ac Stark shift of $m_{g}%
=\pm1$ states relative to the $m_{g}=0$ state then generates an effective
quadratic Zeeman energy shift. On the other hand the atomic population can be
redistributed in the ground state manifold via the two-body $s$-wave spin
exchange collisions, which are described by the numbers $c_{0}=4\pi\hbar
^{2}\left(  2a_{2}+a_{0}\right)  /3m_{a}$ and $c_{2}=4\pi\hbar^{2}\left(
a_{2}-a_{0}\right)  /3m_{a}$ with $m_{a}$ the atom mass and $a_{f}$ the
$s$-wave scattering lengths in the hyperfine channel with a total spin $f=0$
or $2$ \cite{spinor condensate}. We anticipate that this model can be readily
implemented in experiment with the recent advance in coupling ring cavity with
cold atoms \cite{ring cavity1} and BECs \cite{ring cavity2}.

For the present system we apply single-mode approximation (SMA), under which
all three atomic spin states are described by the same spatial wavefunction
$\psi\left(  \mathbf{r}\right)  $. SMA is appropriate for a condensate whose
size is smaller than the spin healing length $\xi_{s}=h/\sqrt{2m_{a}\left\vert
c_{2}\right\vert n}$ ($n$ is the atomic density). The case beyond SMA and with
unbiased driving field was considered in \cite{our model2}.

After adiabatically eliminating the excited atomic level, the atom-cavity
system can be described by the following Hamiltonian in a rotating frame with
$\hbar=1$:%
\begin{equation}
\hat{H}=\hat{H}_{0}+\left[  U_{0}\left(  \hat{c}_{+}^{\dagger}\hat{c}_{+}%
+\hat{c}_{-}^{\dagger}\hat{c}_{-}\right)  -\delta_{c}\right]  \hat{a}%
^{\dagger}\hat{a}+i\varepsilon_{p}\left(  t\right)  \left(  \hat{a}^{\dagger
}-\hat{a}\right)  \label{eq_h}%
\end{equation}
where $U_{0}$ characterizes the strength of atom-photon coupling and
$\delta_{c}=\omega_{p}-\omega_{c}$ is the cavity-pump detuning. $\hat{H}_{0}$
describes the dynamics of the spinor condensate \cite{spinor condensate,
hanPRL1998} and is given by%
\begin{equation}
\hat{H}_{0}=\frac{\lambda}{N}\hat{c}_{a}^{\dagger}\hat{c}_{a^{\prime}%
}^{\dagger}\mathbf{F}_{ab}\cdot\mathbf{F}_{a^{\prime}b^{\prime}}\hat{c}%
_{b}\hat{c}_{b^{\prime}} \label{eq_h0}%
\end{equation}
with $\lambda=Nc_{2}\int d\mathbf{r}\left\vert \psi\left(  \mathbf{r}\right)
\right\vert ^{4}/2$. Here, the total particle number $N=\sum_{s}N_{s}$ is a
constant-of-motion, $\hat{c}_{s}$ ($\hat{c}_{s}^{\dagger}$) is the bosonic
annihilation (creation) operator of the atomic spin-$s$ ($s=0,\pm1$) state,
and the indices $a$, $a^{\prime}$, $b$, $b^{\prime}$ are summed over the
spins. $\mathbf{F}$ are spin-1 matrices with%
\begin{align}
F_{x}  &  =\frac{1}{\sqrt{2}}\left(
\begin{array}
[c]{ccc}%
0 & 1 & 0\\
1 & 0 & 1\\
0 & 1 & 0
\end{array}
\right)  \text{, }F_{y}=\frac{i}{\sqrt{2}}\left(
\begin{array}
[c]{ccc}%
0 & -1 & 0\\
1 & 0 & -1\\
0 & 1 & 0
\end{array}
\right)  ,\nonumber\\
F_{z}  &  =\left(
\begin{array}
[c]{ccc}%
1 & 0 & 0\\
0 & 0 & 0\\
0 & 0 & -1
\end{array}
\right)  . \label{eq_spin matrix}%
\end{align}

The evolution of the cavity-spinor BEC system can be described by the master
equation%
\begin{equation}
\frac{d\hat{\rho}}{dt}=-i\left[  \hat{H},\hat{\rho}\right]  +\kappa\left(
2\hat{a}\hat{\rho}\hat{a}^{\dagger}-\hat{a}^{\dagger}\hat{a}\hat{\rho}%
-\hat{\rho}\hat{a}^{\dagger}\hat{a}\right)  , \label{eq_master}%
\end{equation}
with $\hat{\rho}$ denoting the total density operator for the atomic spin and
cavity degrees of freedom.

The mean-field equations of motion for the $\mathcal{C}$-numbers
$\alpha=\left\langle \hat{a}\right\rangle $ and $\left\langle \hat{c}%
_{s}\right\rangle =\sqrt{N\rho_{s}}\exp\left(  -i\theta_{s}\right)  $
($\rho_{s}$ is the population normalized with respect to the total atom number
$N$ while $\theta_{s}$ is the corresponding phase) can then be derived from
the master equation (\ref{eq_master}) as
\begin{subequations}
\label{eq_mean field}%
\begin{align}
\dot{\alpha}  &  =\left[  i\delta_{c}-iU_{0}N\left(  1-\rho_{0}\right)
-\kappa\right]  \alpha+\varepsilon_{p}\left(  t\right)
,\label{eq_mean field_a}\\
\dot{\rho}_{0}  &  =2\lambda\rho_{0}\sqrt{\left(  1-\rho_{0}\right)
^{2}-m^{2}}\sin\theta,\label{eq_mean field_b}\\
\dot{\theta}  &  =-2U_{0}\left\vert \alpha\right\vert ^{2}+2\lambda\nonumber\\
&  \times\left[  1-2\rho_{0}+\frac{\left(  1-\rho_{0}\right)  \left(
1-2\rho_{0}\right)  -m^{2}}{\sqrt{\left(  1-\rho_{0}\right)  ^{2}-m^{2}}}%
\cos\theta\right]  , \label{eq_mean field_c}%
\end{align}
where $\theta=2\theta_{0}-\theta_{+}-\theta_{-}$ is the relative phase, and
$m=\rho_{+}-\rho_{-}$ the magnetization. Here $\cdot$ means derivative with
respect to time $t$. For simplicity we assume zero magnetization $m=0$ in the
following discussion.

At this point we specify the parameters used in the present work: For spinor
$^{87}$Rb condensate considered in \cite{chapmanNC2016}, $\lambda=-2\pi
\times14$ Hz and $N=4\times10^{4}$, for typical cavity setup we assume that
$\kappa=2\pi\times1$ MHz, $U_{0}=-2\pi\times10$ Hz, $\varepsilon_{0}=4\kappa$
and $f_{0}=0.1$. By considering the fact that the cavity decay rate $\kappa$
is typically much larger than both the frequency of atomic spin oscillation
(characterized by the intrinsic frequency $\lambda$) and the modulation
frequency $\omega_{m}$ (around hundreds Hz as we will show below), we can
adiabatically eliminate $\alpha$ from Eq. (\ref{eq_mean field_a}) and replace
$\alpha$ in Eq. (\ref{eq_mean field_c}) with%
\end{subequations}
\begin{equation}
\alpha\left(  t\right)  \approx\frac{\varepsilon_{p}\left(  t\right)  }%
{\kappa-i\delta_{c}+iU_{0}N\left(  1-\rho_{0}\right)  }.
\label{eq_adiabatical elimination}%
\end{equation}
Thus $\left\vert \alpha\left(  t\right)  \right\vert ^{2}\approx\left\vert
\alpha_{0}\right\vert ^{2}\left[  1+2f_{0}\sin\left(  \omega_{m}t\right)
\right]  $ with $\alpha_{0}=\varepsilon_{0}/\left[  \kappa-i\delta_{c}%
+iU_{0}N\left(  1-\rho_{0}\right)  \right]  $, where we have kept only the
lowest order in $f_{0}$ by considering weak driving.

By introducing $\theta\left(  t\right)  =\phi\left(  t\right)  +z\cos\left(
\omega_{m}t\right)  $ with $z=2\omega_{0}f_{0}/\omega_{m}$ and $\omega
_{0}=2U_{0}\left\vert \alpha_{0}\right\vert ^{2}$, Eqs. (\ref{eq_mean field_b}%
) and (\ref{eq_mean field_c}) become%
\begin{align}
\dot{\rho}_{0}  &  =2\lambda\rho_{0}\left(  1-\rho_{0}\right)  \sum
_{n=-\infty}^{\infty}J_{n}\left(  z\right)  \sin\left[  \phi+n\left(
\omega_{m}t+\frac{\pi}{2}\right)  \right]  ,\nonumber\\
\dot{\phi}  &  =-\omega_{0}+2\lambda\left(  1-2\rho_{0}\right)  \left\{
1\right. \nonumber\\
&  \left.  +\sum_{n=-\infty}^{\infty}J_{n}\left(  z\right)  \cos\left[
\phi+n\left(  \omega_{m}t+\frac{\pi}{2}\right)  \right]  \right\}  ,
\label{eq_tempo}%
\end{align}
where we have implicitly assumed that for evolution at high field with
relatively large $\left\vert \omega_{0}/\lambda\right\vert $ the system is in
the Zeeman energy dominated regime in which the oscillation dynamics are
suppressed, and consequently $\rho_{0}$ and $z$\ can be assumed to be
approximately constant \cite{chapmanNC2016}. Note also that the Jacobi-Anger
expansions%
\begin{align}
\cos\left(  z\cos\varphi\right)   &  =\sum_{n=-\infty}^{\infty}J_{n}\left(
z\right)  \cos\left[  n\left(  \varphi+\frac{\pi}{2}\right)  \right]
,\nonumber\\
\sin\left(  z\cos\varphi\right)   &  =\sum_{n=-\infty}^{\infty}J_{n}\left(
z\right)  \sin\left[  n\left(  \varphi+\frac{\pi}{2}\right)  \right]
\label{eq_jacobi expansion}%
\end{align}
have been used in deriving Eqs. (\ref{eq_tempo}), where $J_{n}\left(
z\right)  $ is the $n$th-order Bessel function of the first kind.

Replacing $\phi\rightarrow\phi-n\left(  \omega_{m}t+\pi/2\right)  $, one can
see that only at some specific values of $n=k$ with $k\omega_{m}\sim\omega
_{0}$ the value of $\phi$ \textit{does not} monotonically depend on $t$, i.e.,
yielding nonzero time average of $\dot{\rho}_{0}$. Around these specific
values of $k$ giving rise to parametric resonances, Eqs. (\ref{eq_tempo})
resort to%
\begin{align}
\dot{\rho}_{0}  &  =2\lambda\eta_{k}\rho_{0}\left(  1-\rho_{0}\right)
\sin\phi,\nonumber\\
\dot{\phi}  &  =\delta_{k}+2\lambda\left(  1-2\rho_{0}\right)  \left(
1+\eta_{k}\cos\phi\right)  , \label{eq_resonance}%
\end{align}
where $\eta_{k}=J_{k}\left(  z\right)  $ and $\delta_{k}=k\omega_{m}%
-\omega_{0}$. The equations-of-motion (\ref{eq_resonance}) have similar form
as the secular equations derived in \cite{gerbier2018}. However one should
notice that $\delta_{k}$ relates to $\left\vert \alpha_{0}\right\vert ^{2}$
and thus is a complex function of $\rho_{0}$, which introduces nonlinearity
into the system.

To illustrate the dynamical properties near parametric resonance, one can use
$\dot{\rho_{0}}=-2\partial H_{k}/\partial\phi$ and $\dot{\phi}=2\partial
H_{k}/\partial\rho_{0}$ to construct, in terms of two conjugate variables
$\rho_{0}$ and $\phi$, the following mean-field Hamiltonian $H_{k}$:%
\begin{equation}
H_{k}=\lambda\rho_{0}\left(  1-\rho_{0}\right)  \left(  1+\eta_{k}\cos
\phi\right)  +U_{k}\left(  \rho_{0}\right)  , \label{eq_secular energy}%
\end{equation}
where%
\begin{equation}
U_{k}\left(  \rho_{0}\right)  =\frac{k\omega_{m}}{2}\rho_{0}+\frac
{\varepsilon_{0}^{2}}{N\kappa}\arctan\left[  \frac{NU_{0}}{\kappa}\left(
1-\rho_{0}\right)  -\frac{\delta_{c}}{\kappa}\right]  \label{eq_potential}%
\end{equation}
represents the cavity-mediated atom-atom interaction.

\section{cavity-amplified parametric resonance}

\label{sec_amplify}

We first consider the cavity-free case where in Eqs. (\ref{eq_mean field})
$U_{0}\left\vert \alpha\right\vert ^{2}$ represents a quadratic Zeeman shift
independent of $\rho_{0}$, then $U_{k}\left(  \rho_{0}\right)  $ in Eq.
(\ref{eq_potential}) resorts to $\delta_{k}\rho_{0}/2$. If the periodic
modulation is not applied ($f_{0}=0$) one can estimate that $\left\vert
U_{0}\left\vert \alpha\right\vert ^{2}/\lambda\right\vert \approx11.4$ at
$\delta_{c}=0$. One can further show that \cite{zhangPRA2005}, under this high
field the maximum oscillation amplitude for $\rho_{0}$ is approximately $0.02$
when $\rho_{0}\left(  0\right)  =0.5$ and goes to zero when $\rho_{0}\left(
0\right)  =0$ or $1$. When one approaches the $k$-th parametric resonance with
the periodic modulation applied, we can make use of Eq.
(\ref{eq_secular energy}) and rewrite Eqs. (\ref{eq_resonance}) as%
\begin{align}
\left(  \dot{\rho}_{0}\right)  ^{2}  &  =4\lambda^{2}\rho_{0}^{2}\left(
1-\rho_{0}\right)  ^{2}\left\{  \eta_{k}^{2}-\left[  \frac{H_{k}\left(
\rho_{0}\left(  0\right)  ,\phi\left(  0\right)  \right)  }{\lambda\rho
_{0}\left(  1-\rho_{0}\right)  }\right.  \right. \nonumber\\
&  \left.  \left.  -\frac{\delta_{k}}{2\lambda\left(  1-\rho_{0}\right)
}-1\right]  ^{2}\right\}  . \label{eq_pollinomial}%
\end{align}
Eq. (\ref{eq_pollinomial}) typically represents undamped cubic anharmonic
oscillator whose analytical solution can generally be written in the form of
Jacobi elliptic functions.\begin{figure}[h]
\includegraphics[width=8cm]{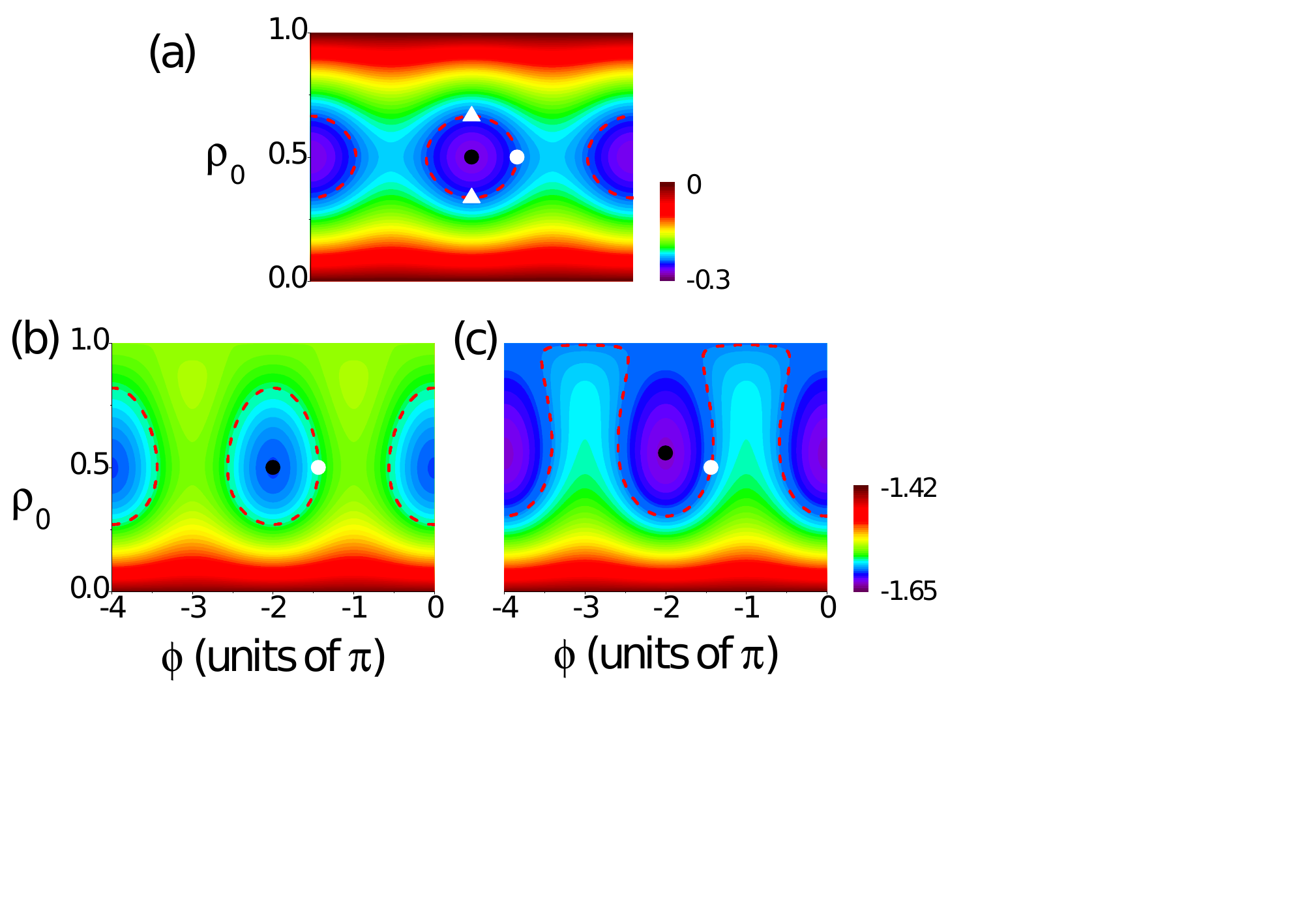}\caption{{\protect\footnotesize Phase-space
contour plot of }$H_{k}${\protect\footnotesize \ (in unit of }$\left\vert
\lambda\right\vert ${\protect\footnotesize ) at }$k=-1$%
{\protect\footnotesize \ resonance. (a) Cavity-free case with }$\delta_{k}%
=0${\protect\footnotesize . The case incoporating cavity backaction are shown
in (b) }$\delta_{k}=0${\protect\footnotesize \ and (c) }$\delta_{k}%
=0.09\lambda${\protect\footnotesize \ with }$\delta_{c}=-0.35\kappa
${\protect\footnotesize . The red-dashed lines refer to the contour determined
by the initial state of the system. The white dots refer to the initial state
of the system, while the black dots refer to equilibrium position and the
white triangles refer to the states when the system passing through the
equilibrium position in the pendulum analogy.}}%
\label{fig_contour}%
\end{figure}

Physical insights into the oscillation properties can be obtained via the
phase-space contour plot of $H_{k}$. We assume that the spinor condensate is
initially prepared in a state with $\rho_{0}\left(  0\right)  =0.5$ and
$\theta\left(  0\right)  =-\pi$ (corresponding to an effective large negative
quadratic Zeeman energy as compared with \cite{chapmanNC2016} in which the
initial state is $\theta\left(  0\right)  =\pi$ with a large positive
quadratic Zeeman energy, $\phi\left(  0\right)  =\theta\left(  0\right)
-z+k\pi/2$). When the driving frequency $\omega_{m}$ is appropriately tuned to
the $k=-1$ resonance with $\delta_{k=-1}=0$, the equal-$H_{-1}$ contour
diagram in the phase space defined by the conjugate pair $\left(  \phi
,\rho_{0}\right)  $ is plotted in Fig. \ref{fig_contour}(a). The contour plot
typically reproduces the phase diagram of a simple pendulum, indicating that
the system evolves along a contour (marked as a red-dashed line) determined by
its initial state (marked as a white dot). The center of the contour (marked
as a black dot) represents the equilibrium position in the pendulum analogy,
which is a stable stationary solution of Eqs. (\ref{eq_resonance}) (the
dynamical properties of the stationary solutions can be studied via the
standard linear stability analysis). The two points marked as white triangles
are two real stationary solutions of Eq. (\ref{eq_pollinomial}) located in the
region $\rho_{0}\in\left[  0,1\right]  $, symbolizing a pendulum passing
through its equilibrium position with maximum speed from different direction.
Their difference is the oscillation amplitude taking the value around $0.33$.

When the cavity backaction is taken into account, one should notice that the
value of $\delta_{k}$ is implicitly $\rho_{0}$-dependent. We first assume that
the driving frequency $\omega_{m}$ is appropriately tuned to $\delta_{k=-1}=0$
with respect to the initial state of $\rho_{0}\left(  0\right)  =0.5$ and
$\theta\left(  0\right)  =-\pi$, and the corresponding phase diagram is shown
in Fig. \ref{fig_contour}(b). Although the contour plot still captures the
main features of a pendulum, its topology changes as compared with Fig.
\ref{fig_contour}(a). In this case one cannot find stationary solutions of Eq.
(\ref{eq_pollinomial}) in the $\rho_{0}\in\left[  0,1\right]  $ region,
implying a non-rigid pendulum. The oscillation amplitude is estimated to take
the value of $0.55$, which is much larger than that of the cavity-free case.
If $\omega_{m}$ is tuned slightly deviate from the resonance with
$\delta_{k=-1}=0.09\lambda$, as shown in Fig. \ref{fig_contour}(c), the
red-dashed line changes its topology from a closed to an open line, and in the
pendulum analogy it signals that the pendulum swings all the way over the
vertical upright position and continues with the same direction of swing. In
this case the oscillation amplitude has the maximum value of about $0.7$,
doubles as compared to the cavity-free case. A drastic topology change is
usually associated with additional fixed points (more than $1$ at $\phi=n\pi
$), which can be determined from the stationary solutions of Eqs.
(\ref{eq_resonance}). From numerical simulations we find that for the $k=-1$
resonance additional fixed points appear in the region $\delta_{c}\in\left[
-0.24\text{, }-0.68\right]  \kappa$ for the present parameter setup,
indicating that one can seek parametric resonance amplification in this
parameter region. \begin{figure}[h]
\includegraphics[width=8cm]{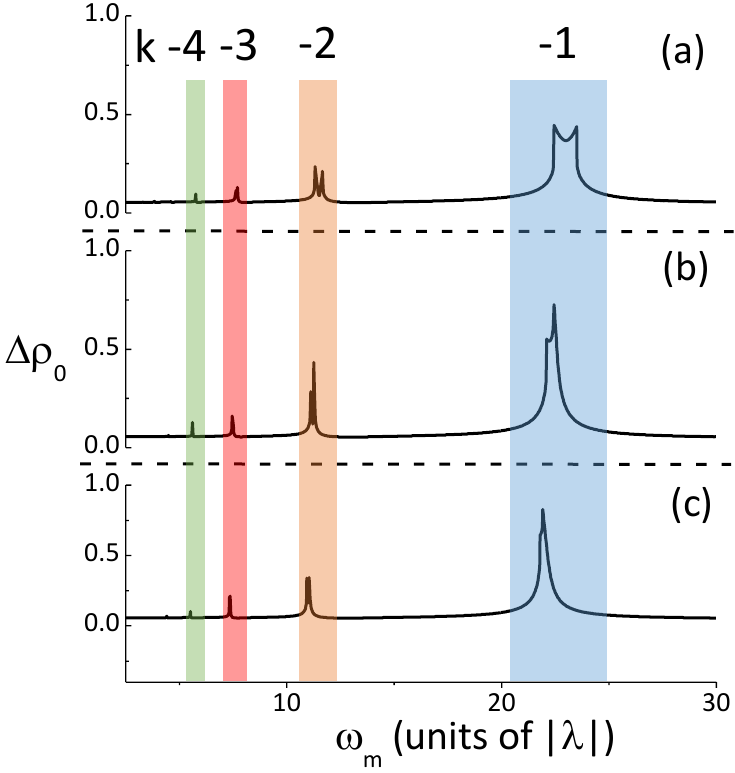}\caption{{\protect\footnotesize Oscillation
amplitude }$\Delta\rho_{0}${\protect\footnotesize \ versus modulation
frequency }$\omega_{m}${\protect\footnotesize \ (in unit of }$\left\vert
\lambda\right\vert ${\protect\footnotesize ) for (a) cavity-free case and the
cases with cavity backaction at (b) }$\delta_{c}=-0.35\kappa$%
{\protect\footnotesize ; (c) }$\delta_{c}=-0.4\kappa${\protect\footnotesize .
The numbered color zone indicates the parametric region in which the }%
$k${\protect\footnotesize -th order resonance is excited.}}%
\label{fig_resonance}%
\end{figure}

A sketch of cavity-mediated parametric resonance is presented in Fig.
\ref{fig_resonance} via numerical simulations of Eqs. (\ref{eq_mean field}),
in which the regions of different $k$-th order resonances (from $k=-1$ to
$-4$) can be well identified. Since $\omega_{0}<0$ (due to $U_{0}<0$), on
parametric resonances $k$ should take negative values. One can notice that the
oscillation amplitude $\Delta\rho_{0}$ significantly decreases for higher
$\left\vert k\right\vert $-th order resonance and those resonances beyond
$k=-4$ are not marked as the oscillation amplitudes are too small to be
unambiguously distinguished from those not excited. This can be traced to the
coupling coefficient $\eta_{k}=J_{k}\left(  z\right)  =J_{k}\left(
2\omega_{0}f_{0}/\omega_{m}\right)  \sim J_{k}\left(  k/5\right)  $, from
which one can estimate that the value of $\eta_{k}$ decays from $10^{-1}$ to
$10^{-4}$ when $k$ varies from $-1$ to $-5$. This indicates that
high-$\left\vert k\right\vert $-th order parametric resonances are much less
likely to be excited. In the pendulum analogy it corresponds to the case that
the system evolves along an ellipse with large curvature, i.e., the pendulum
velocity is small while passing through the equilibrium position.

On resonance the oscillation amplitude $\Delta\rho_{0}$ can display a typical
two-peak structure, as can be seen from the $k=-1$ and $-2$ resonances for the
cavity-free case shown in Fig. \ref{fig_resonance}(a). The exact resonance
point $\omega_{m}=\omega_{0}/k$ locates in the middle of the two peaks, which
is also demonstrated in experiment \cite{gerbier2018}. In experiment
\cite{chapmanNC2016} population $\rho_{0}$ are measured after $100$ ms of
parametric excitation and near the lowest-order resonance population $\rho
_{0}$ behave as a sinusoidal function of $\omega_{m}$ with the resonance point
on the node, which also supports our predictions here. The peaks signal the
critical points at which the pendulum possesses enough energy to pass through
the top position, and they also represent dynamical phase transitions of the
system from $\phi$-running modes to $\phi$-$\pi$ modes. Cavity-induced
nonlinearity substantially modifies the topology of the phase diagram and as
such the two peaks merge into one, as shown in Fig. \ref{fig_resonance}(b) and (c).

More importantly, through cavity-mediated parametric excitation the
oscillation amplitude $\Delta\rho_{0}$ can be significantly amplified. For the
lowest $k=-1$ resonance, Fig. \ref{fig_resonance} demonstrates that cavity
backaction can amplify the oscillation amplitude to the value of $0.83$ as
compared with $0.45$ in the cavity-free case. For high-order resonances such
as $k=-3$, $\Delta\rho_{0}$ can still be amplified to $0.21$ as compared with
the cavity-free value of $0.13$. These results suggest that cavity backaction
can not only make the low-order parametric resonances more prominent, but also
can make the detection of the original weak high-order resonances easier.

\section{measurement discussion}

\label{sec_measure}

In experiments \cite{chapmanNC2016,gerbier2018} spin dynamics are probed via
Stern-Gerlach imaging, which performs fluorescence detection or absorption
imaging after a time-of-flight of spinor condensate in a magnetic field
gradient separating the different spin components. The condensate is
destructed after each detection, which means one will have to repeat the
experiment many times to measure the dynamics. Since the intracavity photon
number $\left\vert \alpha\right\vert ^{2}$ relates to the normalized spin
population $\rho_{0}$ as can be seen from Eq.
(\ref{eq_adiabatical elimination}), this indicates that it can be used for
observing real-time evolution of spin dynamics.\begin{figure}[h]
\includegraphics[width=8cm]{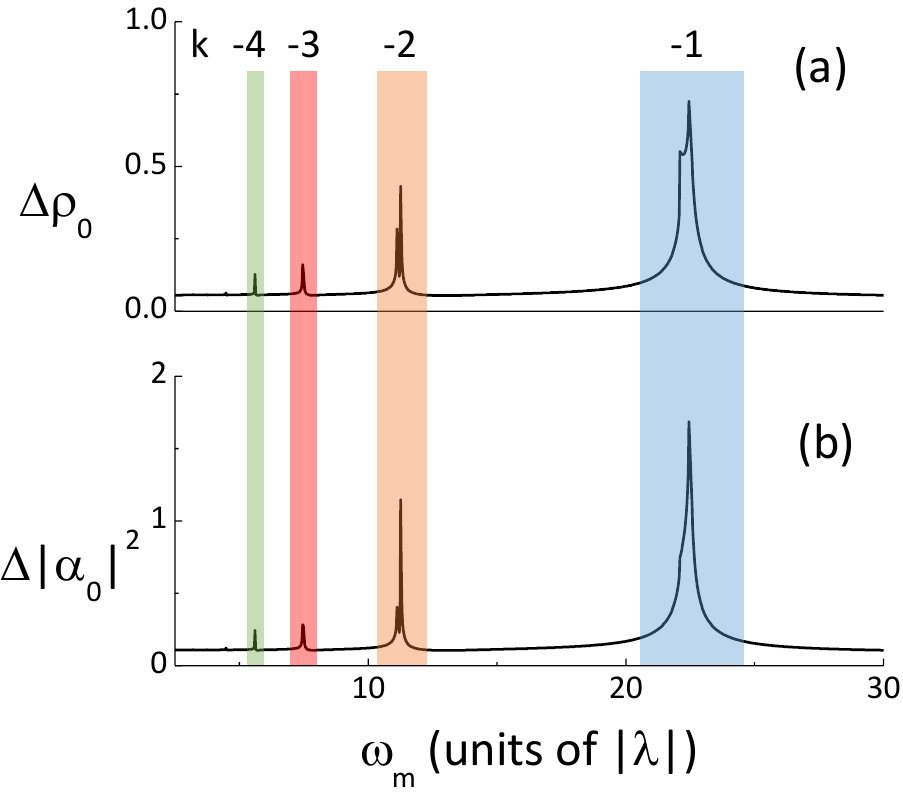}\caption{{\protect\footnotesize (a)
Oscillation amplitude }$\Delta\rho_{0}${\protect\footnotesize \ and (b) the
corresponding cavity oscillation amplitude }$\Delta\left\vert \alpha
_{0}\right\vert ^{2}${\protect\footnotesize \ versus modulation frequency
}$\omega_{m}${\protect\footnotesize \ (in unit of }$\left\vert \lambda
\right\vert ${\protect\footnotesize ) at }$\delta_{c}=-0.35\kappa
${\protect\footnotesize .}}%
\label{fig_measure}%
\end{figure}

As $\left\vert \alpha\left(  t\right)  \right\vert ^{2}\approx\left\vert
\alpha_{0}\right\vert ^{2}\left[  1+2f_{0}\sin\left(  \omega_{m}t\right)
\right]  $, one can integrate $\left\vert \alpha\left(  t\right)  \right\vert
^{2}$ over several periods of modulation to eliminate the high-frequency
oscillation while during this relatively short time (compared to the
oscillation period) the value of $\left\vert \alpha_{0}\right\vert ^{2}$ is
roughly unchanged. In Fig. \ref{fig_measure} we plot the oscillation amplitude
of spin population $\Delta\rho_{0}$ as well as that of averaged intracavity
photon number $\Delta\left\vert \alpha_{0}\right\vert ^{2}$, the results
indicate that continuous observation of spin dynamics can be realized via
measuring the corresponding averaged intracavity photon number $\left\vert
\alpha_{0}\right\vert ^{2}$. Parametric resonances can also be well
identified. We note that the idea of probing spin dynamics with cavity
transmission spectra was also proposed in \cite{zhangPRA2009}.

\section{summary and outlook}

\label{sec_conclusion}

It is interesting to note that bistability in spin-1 condensate was found in
\cite{gerbier2018}, which is aroused by the dissipation of spinor condensate
and hysteresis (usually associated with bistability) was observed for long
evolution times. In the present work we concentrate on relatively short-time
dynamics in which spin relaxation will not play a significant role. However we
would like to note that the interplay between the atomic spin mixing and the
cavity light field can lead to a strong matter-wave nonlinearity and
bistability, which has been demonstrated in previous works \cite{our model,our
model2}. So certainly one can expect that bistability will take place with
parametric excitations here even for short times at appropriate conditions.

In summary we have studied nonlinear Floquet dynamics of spinor condensate in
an optical cavity. Floquet driving leads to parametric resonance while the
cavity-induced nonlinearity makes it amplified. Since the order of observable
resonances is limited by the maximum quadratic Zeeman energy (maximal magnetic
field) achievable \cite{chapmanNC2016,gerbier2018}, thus the scheme propsed in
the present work provide a way to experimentally probe high-order parametric
resonances without the request of increasing the quadratic Zeeman energy.
Feasibility of real-time observation of spin dynamics via cavity output is
also discussed. Other interesting phenomena in this system which can be
modified via the coupling to the cavity, such as quantum spin squeezing
\cite{chapmanNP2012}, entanglement \cite{youexperiment,massonPRL2019,gil} as
well as phase transition \cite{floquet spinor}, will be left for further
investigation. It is also interesting to note that a quite recent work
\cite{clarkNature2019} demonstrated "Floquet polaritons" via the coupling of
Floquet modulated $^{87}$Rb atoms with cavity light modes.

\begin{acknowledgments}
We thank Han Pu and Yongping Zhang for helpful discussions. This work is
supported by National Natural Science Foundation of China (Grants No.
11374003, No. 11574086), the National Key Research and Development Program of
China (Grant No. 2016YFA0302001), and the Science and Technology Commission of
Shanghai Municipality (Grant No. 16DZ2260200).
\end{acknowledgments}

\end{document}